\documentclass[final,3p,times,twocolumn,authoryear,longtitle]{elsarticle}

\usepackage{amssymb}
\usepackage{amsmath}

\usepackage{booktabs,caption}
\usepackage{threeparttable} 
\usepackage[T1]{fontenc}
\usepackage[utf8]{inputenc}
\usepackage{color} 
\usepackage{hyperref}

\usepackage{aas_macros}

\newcommand{\td}[1]{\, \mathrm{d} #1 \,}

\newcommand{\g}{\ensuremath{\gamma}}

\newcommand{\E}[1]{\times 10^{#1}}

\newcommand{\fermi}{\textit{Fermi}-LAT}
\newcommand{\pksf}{PKS\,1510$-$089}
\newcommand{\pkst}{PKS\,2155$-$304}
\newcommand{\fvar}{\ensuremath{F_\text{var}}}
\defcitealias{WZ26}{Paper~I}

\journal{Journal of High Energy Astrophysics}

\begin{document}

\begin{frontmatter}


\title{20 years of monitoring: PKS\,2155$-$304 and PKS\,1510$-$089 in the eyes of Swift and Fermi.\\ II. PKS\,1510$-$089 and comparison} 

\author[lsw,nwu]{Michael Zacharias}
\ead{m.zacharias@lsw.uni-heidelberg.de}
\author[ifj]{Alicja Wierzcholska}
\ead{alicja.wierzcholska@ifj.edu.pl}
\affiliation[lsw]{organization={Landessternwarte, Universit\"{a}t Heidelberg},
            addressline={K\"{o}nigstuhl 12},
            postcode={D-69117},
            postcodesep={},
            city={Heidelberg},
            country={Germany}}
\affiliation[nwu]{organization={Centre for Space Research, North-West University},
            city={Potchefstroom},
            postcodesep={},
            postcode={2520},
            country={South Africa}}
\affiliation[ifj]{organization={Institute of Nuclear Physics, Polish Academy of Sciences},
            addressline={ul. Radzikowskiego 152},
            postcode={31-342},
            postcodesep={},
            city={Krak\'{o}w},
            country={Poland}}

\begin{abstract}
We present a comprehensive, two-decade, multiwavelength variability study of the blazar PKS\,1510$-$089, one of the most prominent and extensively monitored flat-spectrum radio quasars.
Using \textit{Fermi}-LAT $\gamma$-ray data together with \textit{Swift}-XRT and UVOT observations spanning 2005–2024, we trace the long-term evolution of its flux, interband correlations, and spectral behaviour across the optical, X-ray, and $\gamma$-ray bands.
We find that the HE \g-ray and X-ray flux distributions are log-normal, while the optical distributions are compatible with double-log-normal functions. The latter may be due to contributions from the accretion disk. The range of fluxes in a given band, as well as the fractional variability values are in-line with the expectations that high-energy parts of a given spectral component are more variable than low-energy parts.
No obvious cross-correlations exist between the bands over the 20 years of observations.
The X-ray and $\gamma$-ray spectra are variable, but do not show any trend with flux.
These results are suggestive of different zones being active in the jet of PKS\,1510$-$089 at any given time.
In a previous paper, we used the same techniques to study the high-frequency-peaked BL Lac object PKS\,2155$-$304. Both sources follow the aforementioned trend on the energy-dependent variability of the spectral components, as well as the lack of significant cross-correlations between the studied bands.
While PKS\,2155$-$304 exhibits a harder-when-brighter behaviour in its high-energy part of the synchrotron component, no such behaviour could be found in PKS\,1510$-$089.
Both sources show orphan flares, which can seemingly happen in any band.
In summary, the long-term studies of these two sources reveal that the underlying physics is similar in these apparently different source classes, even though variability patterns keep changing and remain unpredictable.

\end{abstract}




\end{frontmatter}



%
%
\section{Introduction}
\pksf\ is a peculiar blazar. As a member of this source class, its relativistic jet emanating from the vicinity of the central supermassive black hole, points towards Earth. The accretion disk is visible as a big blue bump \citep{Malkanetal1986} in the spectral energy distribution (SED), and the broad optical emission lines lead to \pksf's classification as a flat-spectrum radio quasar (FSRQ). Partially owing to its moderate redshift of $z=0.361$, it was the second FSRQ to be detected at very-high-energy (VHE, $E>100\,$GeV) \g\ rays \citep{hess13}, but is at time of writing the only FSRQ with persistent VHE emission \citep{magic18}. 

Apart from the accretion disk and the emission lines, the multiwavelength (MWL) SED shows the two common broad-band non-thermal emission components. The low-energy one peaks in the far-infrared domain and is attributed to synchrotron emission of relativistic electrons. The high-energy component peaks in the MeV domain and is normally assumed to be inverse-Compton emission of the same electrons. Relativistic protons may also directly or indirectly contribute to the high-energy component \citep[see][for a review]{cerruti20}. However, the SED of \pksf\ has always shown complications, which have led to a considerable debate about the nature of its components. While a standard leptonic one-zone model can be made to work, it requires contributions from both the broad-line region (BLR) and the dusty torus (DT) for the seed photon field of the inverse-Compton component \citep[e.g.,][]{barnacka+14,lei+25}. Other authors have argued that the SED along with other constraints requires at least a two-zone approach \citep[e.g.,][]{nalewajko+12,brown13,prince+19,dengjinag24}.

\pksf\ used to be one of the most active blazars exhibiting strong variability in the optical continuum and the lines \citep{Isleretal2015,Rakshit2020,amadorportes+24}, as well as the high-energy (HE, $E>100\,$MeV) \g-ray, and VHE \g-ray domains \citep[e.g.,][]{saito+15,magic17,hess+21}. Remarkably, there was no evidence for significant correlations between the HE and VHE bands \citep{zacharias+19}. Probably, one of the most surprising events took place in July 2021 \citep{hess23}, when after a relatively strong optical flare (and some variability in the \g-ray band) all activity ceased and the optical and HE bands exhibited a severe flux drop. Additionally, the optical polarization also disappeared after reaching a level of up to $20\%$ during the preceding flare \citep{podjed+24}. On the contrary, the X-ray and VHE \g-ray fluxes barely changed. \cite{hess23} interpreted these observations as strong evidence for at least two emission regions with the variable one having vanished.

Here, we analyse the long-term behaviour of \pksf\ by studying 20 years of data from \textit{Swift} and \textit{Fermi}. This unprecedented data set allows us to obtain an overall picture of this blazar instead of the more common individual-events-focused approach. The variability aspects will be studied following the approach of \citet[hereafter \citetalias{WZ26}]{WZ26}, where we studied the long-term properties of the high-frequency peaked BL Lac object (HBL) \pkst. 
This allows us to compare the timing properties of these two objects, which are archetypical members of their respective blazar categories. This is important, as the observational appearance~-- like SED peak positions, presence of thermal radiation fields, FSRQs being considered FRII galaxies, while BL Lac objects are supposedly FRI galaxies, etc.~-- suggests significant differences between these objects. However, while flares are common in both source categories, it is not obvious whether the timing properties are comparable or different. Given that variability is a common feature in most blazars~-- and, hence, jets~-- the timing information can tell us if the jet-intrinsic physics governing the radiation production are similar.
Hence, the comparison of these sources allows us to gain a more fundamental view of blazar jets and their variability.


The paper is organized as follows. The data analysis procedures are described in Sec.~\ref{sec:ana}, while the variability studies are conducted in Sec.~\ref{sec:varia}. In Sec.~\ref{sec:dis}, we discuss the results of \pksf, and compare them to \pkst. We summarize our findings in Sec.~\ref{sec:sum}.

%
%

\section{Data analyis} \label{sec:ana}

\subsection{Fermi-LAT}
The high-energy (HE) $\gamma$-ray observations of \pksf\ obtained with the Large Area Telescope (LAT) onboard the \textit{Fermi} Gamma-ray Space Telescope span the period from August 4, 2008, to August 8, 2024, corresponding to 16 years of continuous monitoring of the blazar. The data analysis was conducted using the \textit{Fermi} Science Tools (version 2.2.0), the fermipy package (version 1.2.2), and the instrument response functions \verb|P8R3_SOURCE_V3|.

Events were selected within a $15^\circ$ radius around the source position, applying standard quality criteria, including a zenith angle cut of $<90^\circ$ and the \verb|DATA_QUAL|~$>0$ filter to exclude intervals of poor data quality. The analysis was restricted to the 100\,MeV–500\,GeV energy range.

The background model included all sources from the 4FGL-DR3 catalogue within $12^\circ$ of the target. 
For sources located within $3^\circ$, the spectral parameters were left free during the likelihood fitting. The Galactic and isotropic diffuse emissions were modelled using the standard templates \verb|gll_iem_v07.fits| and \verb|iso_P8R3_SOURCE_V3_v1.txt|, respectively.

The source flux and spectral parameters were derived using a binned likelihood analysis with 10 energy bins per decade and spatial bins of $0.1^\circ$. 
Additionally, four new point-like sources~-- J1513.3-0753, J1522.4-0517, J1537.1-1012 and J1533.1-0309~-- were added to the model. They were identified by the method \verb|find_sources| from the fermipy package, requiring a test statistic $(\mathrm{TS}) > 5$.

We construct a light curve with $4\,$d binning. The spectral reconstruction per time bin is done using both a power-law model,

\begin{align}
    \frac{\td{F}}{\td{E}} = N_0 \left( \frac{E}{E_0} \right)^{-\Gamma}
    \label{eq:powerlaw},
\end{align}
and a log-parabola model,

\begin{align}
    \frac{\td{F}}{\td{E}} = N_0 \left( \frac{E}{E_0} \right)^{-\alpha-\beta\log(E/E_0)}
    \label{eq:logparabola},
\end{align}
with normalization $N_0$ at fixed reference energy $E_0= 1\,$GeV, the spectral indices of the power law, $\Gamma$, and of the log parabola, $\alpha$, and the curvature $\beta$.
The $\log$ denotes the natural logarithm. If the difference in test statistics of a given bin between the two reconstructions is $TS_{\rm LP}-TS_{\rm PWL}>9$, the log-parabola result is used, while else the power-law reconstruction is taken.

In the following, only light curve bins fulfilling these criteria are used: $TS>9$, and $F/\Delta F >1.0$, where $F$ and $\Delta F$ are the integrated flux of a time bin and its uncertainty, respectively. These criteria remove insignificant flux bins. 
The combined light curve is shown in Fig.~\ref{fig:1510_mwl_lc}(a), while the evolution of the spectral parameters is shown in panel (b) of the same figure. The weighted average\footnote{The weights are determined from the statistical error of the observed quantity.} flux is $(3.20\pm0.02)\E{-7}\,$ph/cm$^2$/s. The weighted average index (both spectral reconstructions combined\footnote{The individual weighted averages of $\Gamma$ and $\alpha$ agree within errors with each other ($\overline{\Gamma}=2.308\pm0.004$ and $\overline{\alpha}=2.311\pm0.007$) and the combined average. They are also both strongly variable.}) and curvature are $2.309\pm0.004$ and $0.18\pm0.02$, respectively. The flux and index are strongly variable, while the curvature is not.

\subsection{Swift-XRT and Swift-UVOT}
Observations performed between 2005 and 2024, corresponding to ObsIDs 00030797001-00031173234, were processed using the HEASoft software package (v6.35.1). All event files were calibrated and cleaned with the \verb|xrtpipeline| task. The analysis included data from both Photon Counting (PC) and Windowed Timing (WT) modes in the 0.3–10\,keV energy range. For spectral fitting, spectra were grouped with the \verb|grppha| tool to ensure a minimum of 20 counts per bin. The spectral modelling was carried out using a log-parabola function, Eq.~\eqref{eq:logparabola}, and a Galactic hydrogen column density value of $7.13 \times 10^{20}$\,cm$^{-2}$ \citep{HI4PI}, using the \verb|XSPEC| package \citep{Arnaud96}. In all fits, N$_H$ was kept frozen.

\begin{figure*}
\centering
\includegraphics[width=0.98\textwidth]{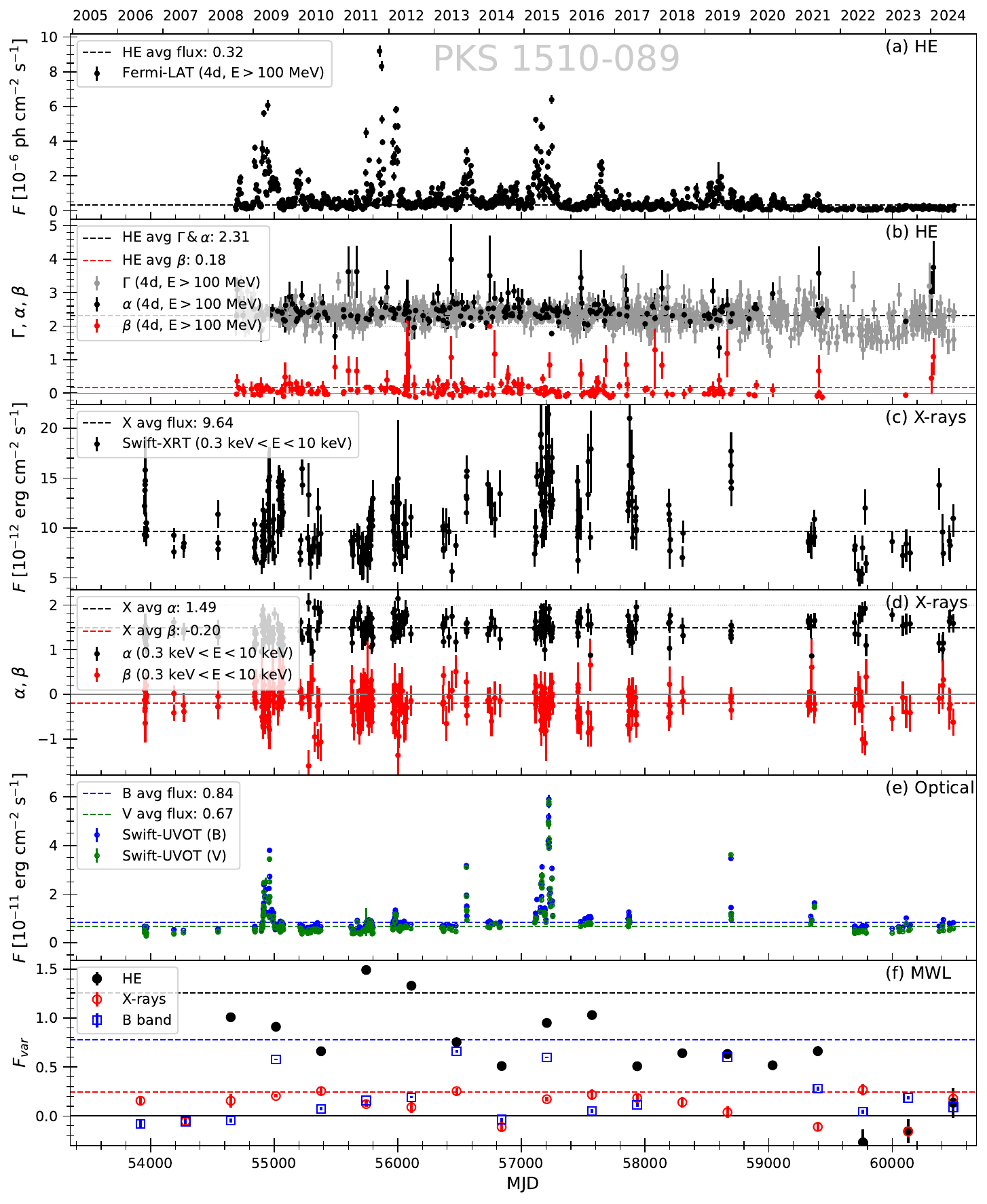}
\caption{Time-evolution of \pksf\ from 2005 to 2024. 
(a) HE \g-ray light curve from \fermi\ combining both power-law and log-parabola spectral reconstructions. The black dashed line marks the weighted average flux of the data shown. 
(b) HE \g-ray spectral parameters with gray points for $\Gamma$, and black and red symbols for $\alpha$ and $\beta$, respectively, in case the log-parabola reconstruction is preferred. The black dashed line gives the weighted average of $\Gamma$ and $\alpha$ combined, while the red dashed line indicates the weighted average of $\beta$. The gray solid and dotted lines mark values of $0.0$ and $2.0$ respectively.
(c) X-ray flux light curve from \textit{Swift}-XRT with the black dashed line giving the weighted average.
(d) X-ray spectral parameters, $\alpha$ and $\beta$, as indicated with the corresponding dashed lines giving the respective weighted averages.
(e) B and V band light curves from \textit{Swift}-UVOT with coloured dashed lines indicating the respective weighted average.
(f) Year-wise \fvar\ values for the four bands shown above. The coloured dashed lines indicate the respective \fvar\ values considering all fluxes of a given band. 
}
\label{fig:1510_mwl_lc}
\end{figure*}

The X-ray light curve is plotted in Fig.~\ref{fig:1510_mwl_lc}(c) with the spectral parameter evolution presented in panel (d). The weighted average of the flux is $(9.64\pm0.07)\E{-12}\,$erg/cm$^{2}$/s, while it is $1.49\pm 0.01$ and $-0.20\pm0.02$ for the index and curvature, respectively. The flux and index are significantly variable, while the curvature is not.

Simultaneously, the source was observed with the UVOT instrument onboard \textit{Swift} in the U (345 nm), B (439 nm), and V (544 nm) filters. For each observation associated with the ObsIDs listed above, instrumental magnitudes were derived using \verb|uvotsource|, with source counts extracted from a circular region of radius 5''. The background was estimated from a nearby, source-free circular region with a 10'' radius. Flux conversion factors were taken from \cite{Poole08}. All UVOT measurements were corrected for Galactic extinction using a reddening value of $E(B-V) = 0.0869$ \citep{schlafly}. Extinction corrections for the individual filters were applied using the $A_{\lambda}/E(B-V)$ coefficients from \citep{Giommi06}.

The B and V band light curves are shown in Fig.~\ref{fig:1510_mwl_lc}(e). The weighted average fluxes are $(8.36\pm0.03)\E{-12}\,$erg/cm$^2$/s and $(6.71\pm0.04)\E{-11}\,$erg/cm$^2$/s for the B and V band, respectively. Both light curves are strongly variable.

%
%

\section{Variability study} \label{sec:varia}

We follow the same procedures as in \citetalias{WZ26} and thus refer to that paper for the details of the methods. We focus on the aspects of interband correlations, the flux distributions, and the fractional variability.

\subsection{Correlations} \label{sec:cor}
\begin{figure*}
\centering
\includegraphics[width=0.98\textwidth]{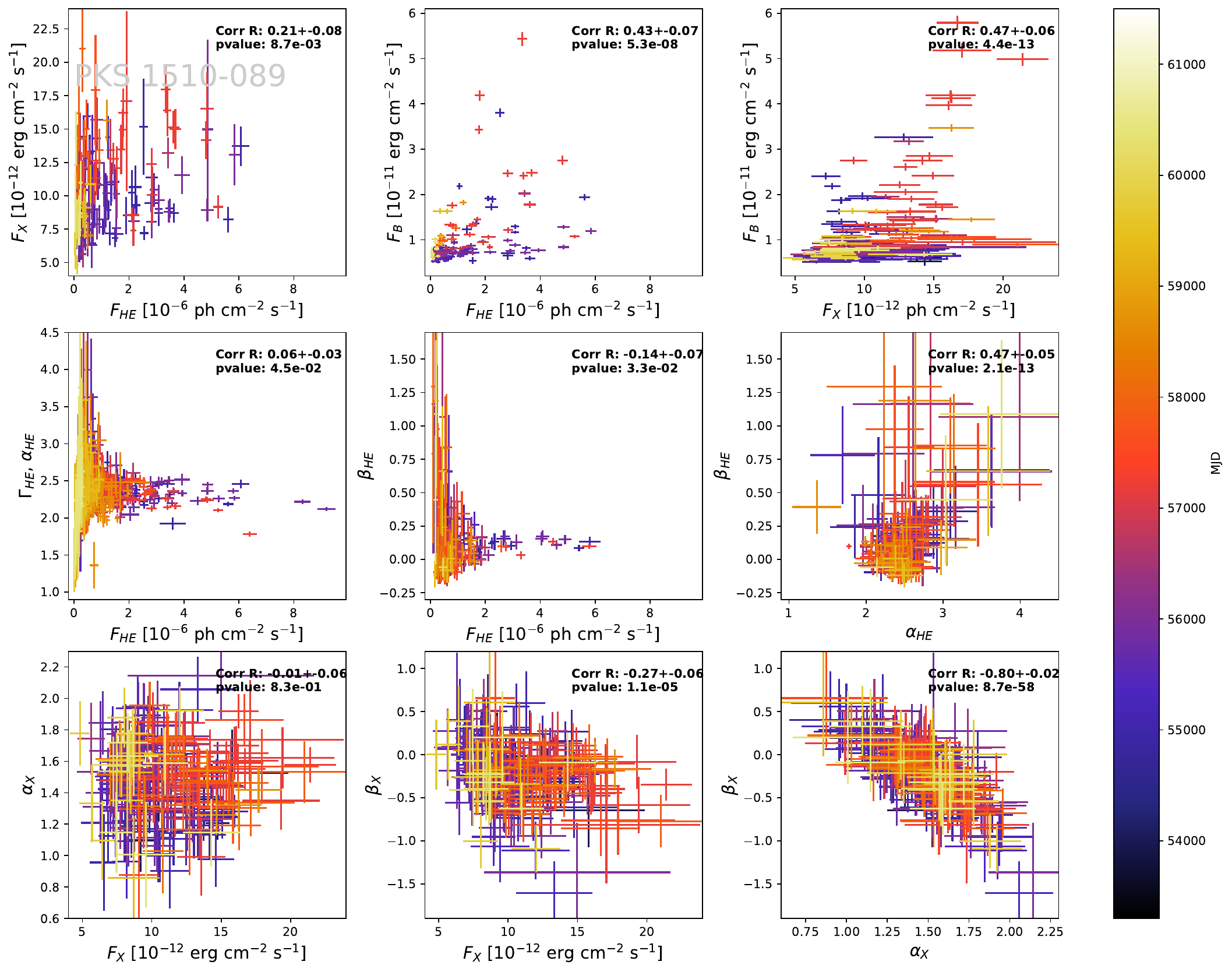}
\caption{(Top row) Correlation plots of flux vs flux for X-rays vs HE \g\ rays (left), B band vs HE \g\ rays (middle), and B band vs X-rays (right).
(2nd row) HE \g-ray index vs flux (left), curvature vs flux (middle) and curvature vs index.
(Bottom row) Same as second row, but for the X-ray data.
In each panel, the Person R coefficient for the correlation is given, as well as the p-value of the probability that a similar or higher R coefficient could be obtained from an uncorrelated data set.
}
\label{fig:1510_flux_flux}
\end{figure*}
\begin{figure*}
\centering
\includegraphics[width=0.98\textwidth]{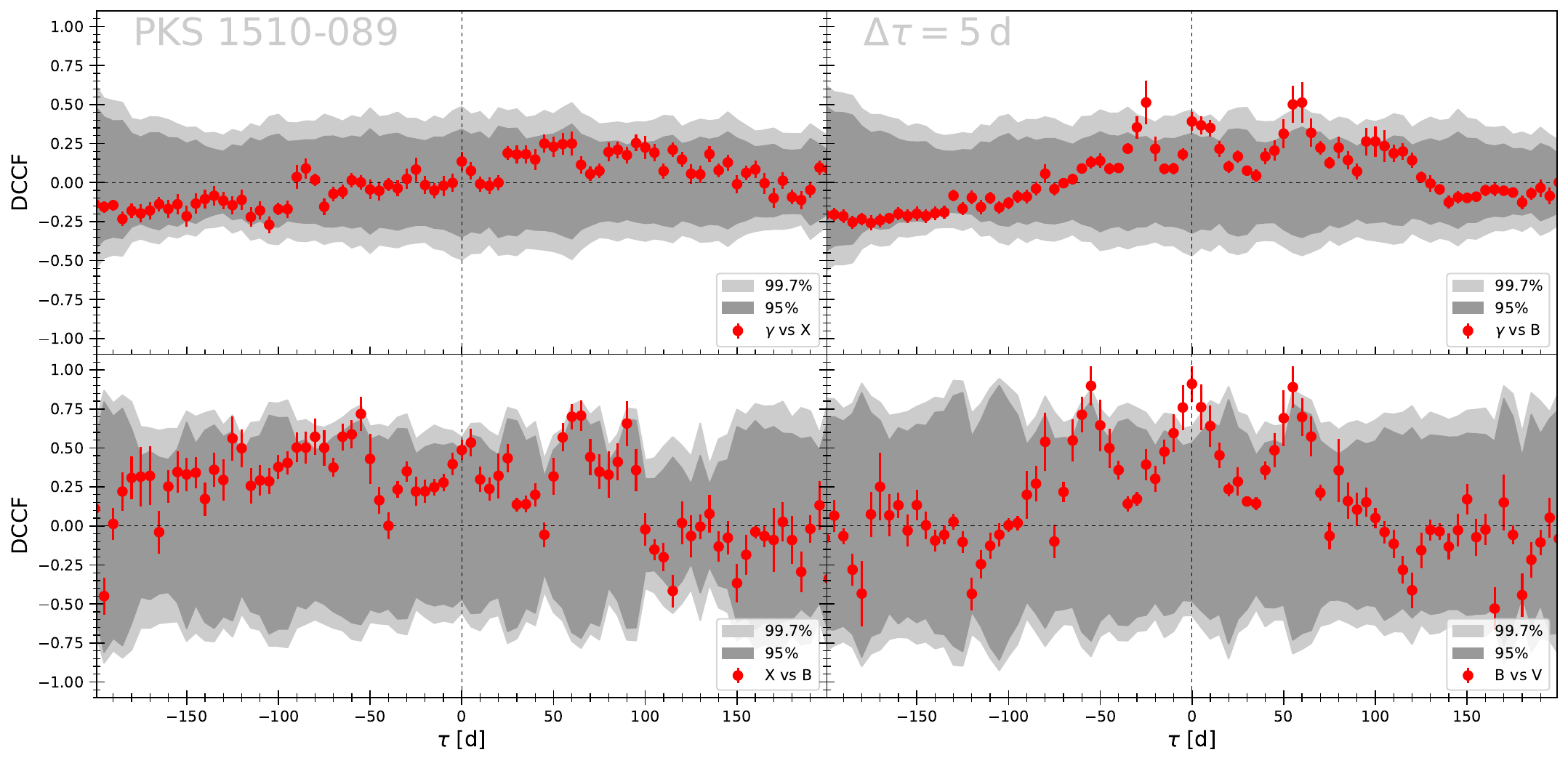}
\caption{DCCF with a time resolution of $\Delta\tau=5\,$d of \pksf\ between the various bands as labelled. The gray bands mark the confidence intervals as indicated.
}
\label{fig:1510_dccf}
\end{figure*}
\begin{table}
\caption{Spectral details of the hard HE spectra. 
Entries below the horizontal line are observations after the flux drop in July 2021.
}
\label{tab:1510_HE_hard_spectra}
\begin{tabular}{lccc}
\hline
MJD &	$F_{\mathrm{HE}}$	&	$\Gamma_{\mathrm{HE}}$, $\alpha_{\mathrm{HE}}$	&	$\beta_{\mathrm{HE}}$	\\
    & $[10^{-7}\,\text{ph/cm}^2\text{/s}]$ & & \\
\hline
55252.7 &	$0.85\pm 0.49$ &	$1.74\pm 0.25$ &	 \\
57244.7 &	$57.83\pm 0.31$ &	$0.99\pm 0.92$ &	$0.24\pm 0.27$ \\ 
57744.7 &	$0.84\pm 0.49$ &	$1.73\pm 0.24$ &	 \\
58048.7 &	$0.82\pm 0.43$ &	$1.66\pm 0.22$ &	 \\
58600.7 &	$3.68\pm 0.50$ &	$1.56\pm 0.45$ &	$0.09\pm 0.14$ \\ 
58768.7 &	$0.47\pm 0.35$ &	$1.56\pm 0.30$ &	 \\
58828.7 &	$1.66\pm 0.64$ &	$1.79\pm 0.19$ &	 \\
59000.7 &	$0.98\pm 0.61$ &	$1.40\pm 0.24$ &	 \\
59012.7 &	$0.50\pm 0.40$ &	$1.38\pm 0.31$ &	 \\
59180.7 &	$0.58\pm 0.45$ &	$1.64\pm 0.31$ &	 \\ 
\hline
59424.7 &	$0.52\pm 0.40$ &	$1.58\pm 0.29$ &	 \\
59484.7 &	$0.17\pm 0.15$ &	$1.50\pm 0.36$ &	 \\
59488.7 &	$0.48\pm 0.25$ &	$1.51\pm 0.21$ &	 \\
59548.7 &	$0.46\pm 0.32$ &	$1.50\pm 0.28$ &	 \\
59632.7 &	$0.52\pm 0.30$ &	$1.74\pm 0.25$ &	 \\
59672.7 &	$0.79\pm 0.44$ &	$1.75\pm 0.25$ &	 \\
59764.7 &	$0.16\pm 0.12$ &	$1.30\pm 0.30$ &	 \\
59804.7 &	$0.58\pm 0.29$ &	$1.64\pm 0.22$ &	 \\
59824.7 &	$0.44\pm 0.31$ &	$1.59\pm 0.28$ &	 \\
59840.7 &	$0.42\pm 0.25$ &	$1.51\pm 0.24$ &	 \\
59876.7 &	$0.43\pm 0.36$ &	$1.60\pm 0.33$ &	 \\
59908.7 &	$1.3\pm 1.1$ &	$1.54\pm 0.35$ &	 \\
59956.7 &	$0.34\pm 0.21$ &	$1.53\pm 0.24$ &	 \\
60008.7 &	$0.30\pm 0.24$ &	$1.60\pm 0.32$ &	 \\
60044.7 &	$0.55\pm 0.43$ &	$1.67\pm 0.31$ &	 \\
60048.7 &	$0.60\pm 0.47$ &	$1.67\pm 0.32$ &	 \\
60168.7 &	$0.81\pm 0.33$ &	$1.75\pm 0.19$ &	 \\
60452.7 &	$0.25\pm 0.24$ &	$1.54\pm 0.38$ &	 \\
60496.7 &	$0.34\pm 0.31$ &	$1.60\pm 0.34$ &	 \\ 
\hline
\end{tabular}
\end{table}
%
%
%
%

Direct correlations between fluxes of various energy bands are studied easiest by plotting the fluxes against each other. This is shown in the top row of Fig.~\ref{fig:1510_flux_flux}. The X-ray-vs-HE panel (top left) does not indicate any obvious correlation. Interestingly, it is indicative of X-ray high states without HE high-states, while there seems to be a minimum level of X-ray flux rise in connection to HE activity. The latter is visible as the white space close to the x-axis. In other words, HE activity is always accompanied by at least a mild reaction in the X-ray domain.

The B-vs-HE panel (top centre) suggests more than one track of correlation. Hence, activity in one band is met with activity in the other, but always in a different manner. The B-vs-X-ray panel (top right) indicates two tracks, of which one also implies a near-orphan X-ray activity. However, given that there is also quite some space filled with points, the correlations are far from clear. For both panels, these statements are underlined by the Pearson R coefficient, which is significant according to its p-value, but the R value itself does not indicate neither high correlation nor none.

With the help of our discrete cross-correlation function (DCCF) routine (for details see \citealt{Taylor+26}, and \citetalias{WZ26}), we study if there may be delayed correlations between these bands. The results are shown in Fig.~\ref{fig:1510_dccf}. There is no correlation between the HE and X-ray light curves, in-line with our conclusions from Fig.~\ref{fig:1510_flux_flux}. In the HE-vs-B case, there are three visible peaks at time delay $\tau\sim-30\,$d, $\tau\sim0\,$d, and $\tau\sim50\,$d. The first one of these is the highest peak with a correlation coefficient of roughly $0.5$; but it is compatible with the $3\sigma$ band and thus not fully significant. While there are some peaks in the X-vs-B case, all of them are well within the $3\sigma$ band, and thus not significant either. 

We also show the B-vs-V case. Given how close these bands are in energy, they represent almost an auto-correlation function. Indeed, the peak at $\tau=0\,$d is highly significant and the DCCF function appears almost axis-symmetric. The decrease from the peak at $\tau=0\,$d suggests a light-curve-intrinsic time scale of roughly $30\,$d. There are also peaks at $\tau\sim\pm60\,$d, which however are not significant.

We also study the spectral evolution of the HE and X-ray bands as shown in Fig.~\ref{fig:1510_flux_flux}. There is no obvious correlation of the HE spectral parameters~-- neither with respect to flux nor among each other, even though weighted constants are ruled out by more than $3\sigma$ in all three cases suggesting significant scatter. While the photon index, $\Gamma$ or $\alpha$, can decrease considerably to values below $2.0$, this is not related to the flux state or the curvature. Within the limits of our data set (esp. considering the $4\,$d binning), there are 29 instances where the index is significantly below $2.0$; that is, even within the range of the 1$\sigma$ statistical error, the value remains below $2.0$. These are listed in Tab.~\ref{tab:1510_HE_hard_spectra}. 
Interestingly, almost two thirds of these "hard states" are attained since July 2021 ($\sim$MJD~59423), when a significant flux drop took place in both the HE and optical light curves \citep{hess23}.
Focusing on the three brightest instances: The first one occurred in August 2015 (MJD~57244.7) during the brightest HE flare of that year. The second and third took place in April (MJD~58600.7) and December (MJD~58828.7) 2019, respectively. The first and second case are the only ones in Tab.~\ref{tab:1510_HE_hard_spectra} reproduced with a log-parabola model, even though $\beta$ does not indicate significant curvature.
Interestingly, neither the HE \g-ray flare in May 2015, where VHE \g-ray variability was detected \citep{magic17}, nor the periods of the first and second major VHE \g-ray flares in May 2016 \citep{hess+21} and July 2019 \citep{hess19ATel} are present in this list despite the occurrence of a very hard HE spectrum especially during the May 2016 event.
This is probably a consequence of the 4\,d binning of the light curve.

In the X-ray domain, there also are no strong correlations (judging by the values of R) between the spectral parameters and the flux. The photon index is, however, strongly variable\footnote{We point out that R and a weighted constant differently probe the data and that the derivation of R does not consider the statistical uncertainties of the data. This can lead to seemingly contradictory results.}, as a weighted constant is rejected with almost $8\sigma$. The curvature is compatible with a weighted constant. Index and curvature seem to be significantly correlated according to R. However, a weighted constant cannot be ruled out, either, as the rejection significance is only $2.2\sigma$. 

\subsection{Flux distribution} \label{sec:fluxdist}
\begin{figure*}
%
\begin{minipage}{0.73\textwidth}
\includegraphics[width=0.98\textwidth]{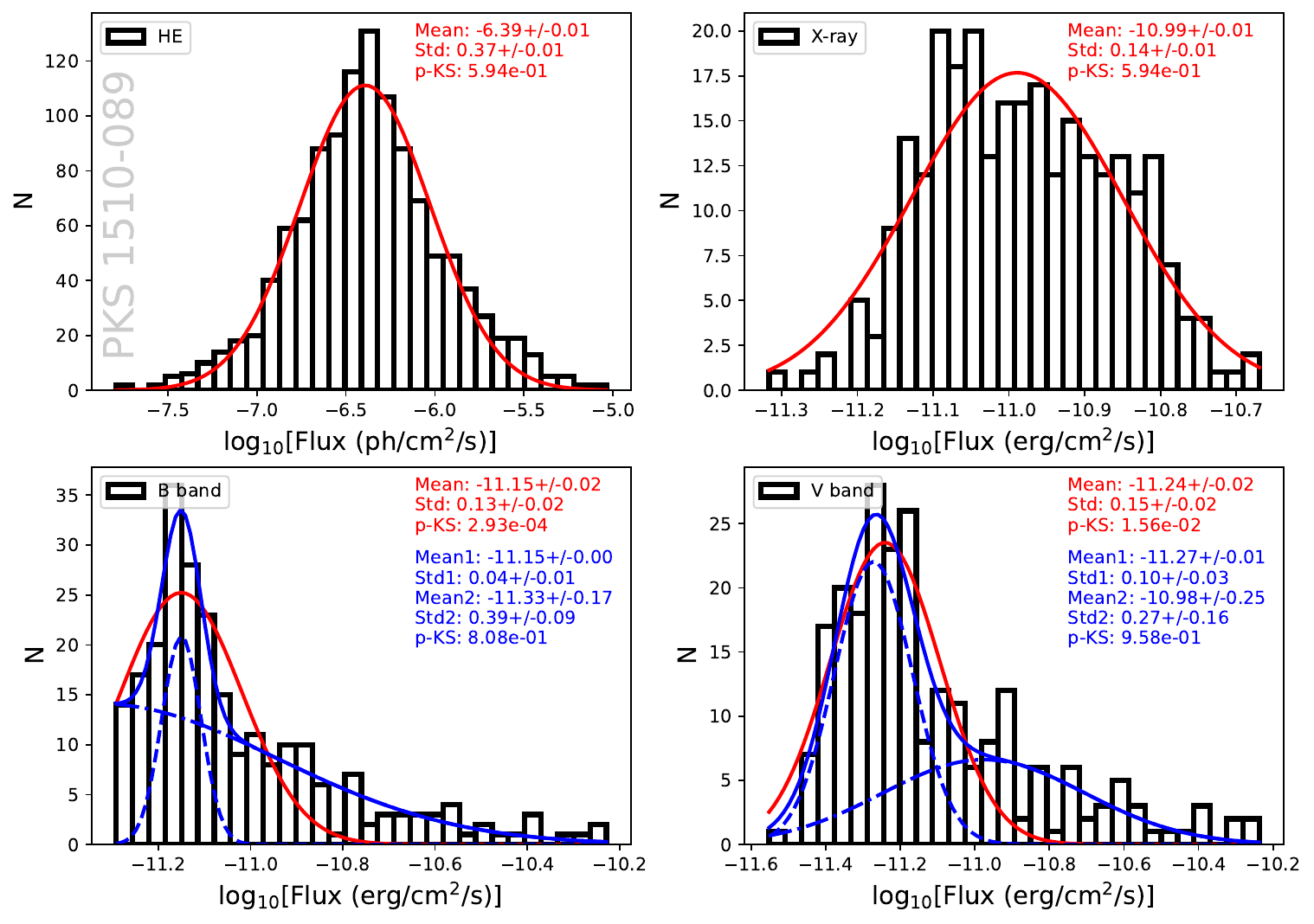}
\end{minipage}
~
\begin{minipage}{0.25\textwidth}
\caption{Flux distributions of the HE \g-ray (top left), X-ray (top right), B band (bottom left) and V band (bottom right) data using logarithmic binning. The red lines mark Gaussian fits with its mean, standard deviation and p-value of the KS test indicated as red text. 
The solid blue line marks a double-Gaussian fit, while the dashed and dash-dotted lines represent the individual components. The corresponding parameter sets are given in blue text, where parameters 1 correspond to the dashed line and parameters 2 to the dash-dotted line.
}
\label{fig:1510_flux_dist}
\end{minipage}
\end{figure*}
\begin{figure*}
\begin{minipage}{0.73\textwidth}
\includegraphics[width=0.98\textwidth]{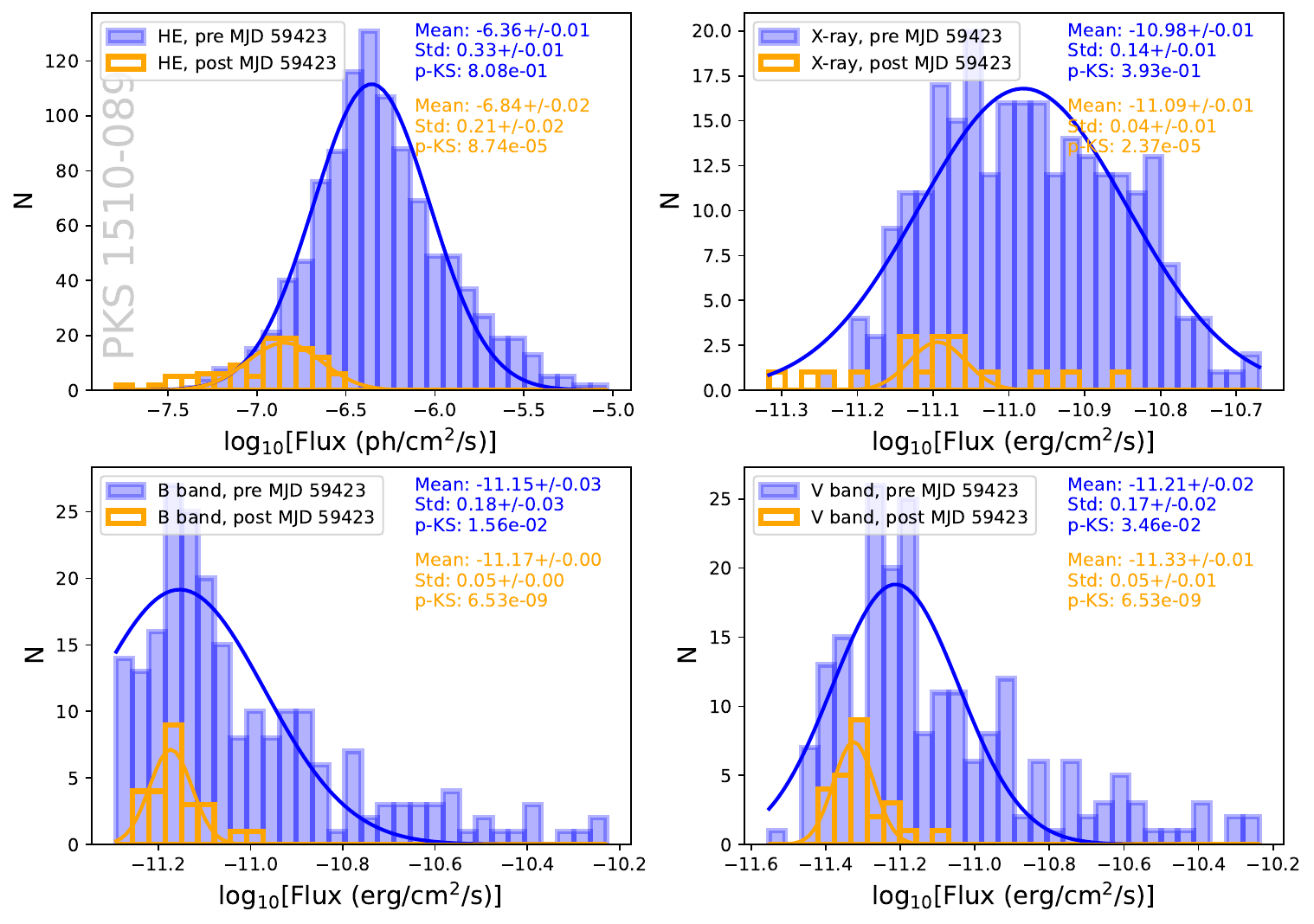}
\end{minipage}
~
\begin{minipage}{0.25\textwidth}
\caption{Similar to Fig.~\ref{fig:1510_flux_dist}, but the blue shaded histogram is for data taken before MJD~59423, while the golden histograms are for data taken afterwards (as in Fig.~\ref{fig:1510_flux_dist}. The blue and golden lines are Gaussian fits to the respective histograms with parameters given in the appropriate colour.
}
\label{fig:1510_flux_dist2}
\end{minipage}
\end{figure*}
%
%
%
The flux distributions are presented in Fig.~\ref{fig:1510_flux_dist} with logarithmic binning. Both the HE and X-ray distributions are compatible with a log-normal distribution. However, both optical bands are not. The goodness-of-fit has been derived with a Kolmogorov-Smirnoff (KS) test with the corresponding p-value given in each panel. We consider the fit as compatible for $p_{\rm KS}>0.05$.

Fitting the optical distributions with two Gaussian functions -- thus, producing a double-log-normal fit -- results in good fits. 
In both cases, the sub-distributions are well separated. Sub-distribution 1 is located around the median flux of the total distribution, is relatively narrow, and thus accounts very well for the narrow peak of the total distribution. Considering the standard deviations ("Std") of these sub-distributions, their respective means are compatible. 
In the B-band, the maximum of sub-distribution 2 is located at the minimum flux of the total distribution, has a relatively low amplitude and extends to the highest fluxes. In the V-band, it peaks at a relatively high flux, but also with low amplitude while being relatively broad.  
The clean separation between the two distributions in both light curves suggests that they belong to different source states. The broadness of sub-distributions 2 suggest highly variable fluxes in this state, while for sub-distributions 1 variability is much less pronounced. It is known that \pksf\ has big blue bump in its SED, which is typically interpreted as a sign of the accretion disk. Thus, it is plausible that state 1 relates to a disk-dominated state, while state 2 is the jet-dominated state.

The range of fluxes attained depends strongly on the energy domain. The HE distribution covers more than 2.5 orders of magnitude, the X-ray domain 0.6 orders of magnitude, and the optical bands up to 1.3 orders of magnitude. If we believe in the separation of the disk- and jet-dominance to the optical distribution, then the optical sub-distribution 2 may only show the high-flux portion of the jet. In this case, the jet flux range could be much larger~-- perhaps in-line with the range of fluxes in the HE band. In any case, the higher the energy band in a spectral component, the broader the flux range.

Furthermore in Fig.~\ref{fig:1510_flux_dist2}, we study the influence of the flux drop that took place in July 2021 \citep{hess23}. In this case, the data is separated by date with data taken before the flux drop in blue shades and the data taken past the flux drop\footnote{We use MJD~59423 for the separation. This is 10 days after the HE flux drop \citep{hess23} in order to avoid any contamination with the evolution of the actual decay.} in gold. 

The Gaussian fits to the blue data agree well with the single-Gaussian fits in Fig.~\ref{fig:1510_flux_dist} judging by the positions of the means being within one standard deviation of each other for corresponding bands.
Indeed, the flux drop mostly influences the HE and optical bands, where the range of fluxes has shrunk considerably to about 1 order of magnitude in the HE band and about 0.3 orders of magnitude in the optical bands. The respective distributions reside at the lower side of fluxes, even though, interestingly, the optical flux attained lower values in the past. 

The Gaussian fits to the golden histograms support this description. The HE \g-ray distribution has shifted to a lower mean and the standard deviation has reduced compared to the blue fit. In the optical bands, the golden fits agree well with the sub-distributions 1 in Fig.~\ref{fig:1510_flux_dist}. This gives additional credit to the aforementioned interpretation that these sub-distributions describe the accretion-disk-dominated state of the total flux distribution, as we know that post flux drop the optical data is well described with the accretion disk only \citep{hess23,podjed+24}.

The after-flux-drop X-ray distribution indicates that the flux drop in the HE and optical band had no influence here. While the lowest fluxes have been attained since this date, the overall distribution is almost as broad as the entire distribution. Unfortunately, the low number of pointings with \textit{Swift} and the spread of the data result in a bad fit of the golden data making an interpretation tricky.


It is, of course, tempting to try to remove the accretion disk flux from the optical fluxes in order to get the true jet fluxes. However, the flux distribution 
suggests that the disk is most likely not constant, either, as the overall minimum optical flux was not reached during the last few years when the jet contribution was low or possibly even absent given the absence of any optical polarization \citep[][]{hess23,podjed+24}. Hence, just using the minimum flux to correct for the accretion disk would not help, as there would still be an unknown amount of ``accretion disk residual flux''.

\subsection{Fractional variability} \label{sec:fracvar}
\begin{figure*}
\centering
\includegraphics[width=0.98\textwidth]{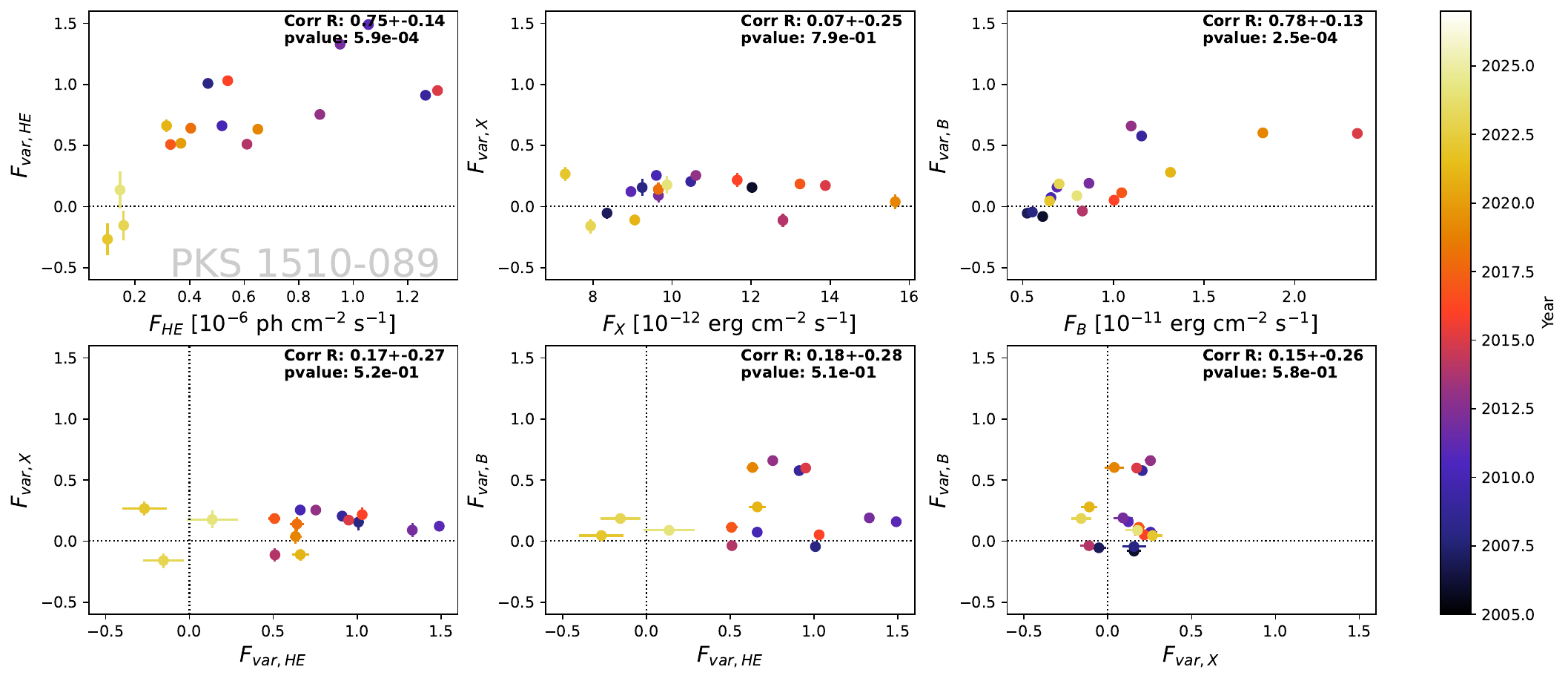}
\caption{(Top row) Year-wise \fvar\ 
as a function of yearly average flux for HE \g\ rays (left), X-ray (middle) and the B band (right). Horizontal error bars mark the average error in a given year. 
(Bottom row) Year-wise \fvar\ vs \fvar\ for X-rays vs HE \g\ rays (left), B band vs HE \g\ rays (middle) and B band vs X-rays (right).
In each panel, the Pearson R coefficient and its p-value are given as in Fig.~\ref{fig:1510_flux_flux}. 
Gray dotted lines mark $\fvar=0$.
}
\label{fig:1510_fvar}
\end{figure*}
%
%
%
%

As described in detail in \citetalias{WZ26}, the fractional variability, \fvar, is a measure to describe the magnitude of variability in a light curve \citep{vaughan+03,poutanen+08,schleicher+19}. In cases without statistically significant variability, the mean-squared flux error, $\overline{\sigma_{\rm err}^2}$, is larger than the variance, $S^2$, and the \fvar\ becomes imaginary. However, in order to keep these values, we calculate $|S^2-\overline{\sigma_{\rm err}^2}|$ and define \fvar\ as negative. We do not calculate an \fvar\ value, if there are less than 3 observations in a year for a given band. 

The \fvar\ by energy band and year is plotted in Fig.~\ref{fig:1510_mwl_lc}(f).
It indicates the trend already seen in the flux distributions: The HE band is more variable than the optical band, which in turn is more variable than the X-ray band. The latter is always $\fvar\lesssim0.3$, while in the HE band $\fvar\gtrsim0.4$ until 2021. Indeed, since the flux drop in July 2021, the variability in The HE domain is at a low level or even absent. 
The optical band attaints any variability up to $\fvar\sim 0.8$. Curiously, during low flux states, the X-ray band can exhibit higher variability than the optical band. This may be related, again, to the accretion disk providing a minimum flux level during low(er) states of the jet.

We also study the correlations of the \fvar\ with the average fluxes in a given year, as well as among the different \fvar. The results are shown in Fig.~\ref{fig:1510_fvar}. The HE and optical band indicate a correlation between the \fvar\ and the average fluxes, while there is none in the X-ray band. 
There is no correlation between the \fvar\ values of the various bands. Hence, strong variability in one band does not necessarily imply strong variations in another. This is in-line with the findings in Sec.~\ref{sec:cor}.

%
%
\section{Discussion} \label{sec:dis}

\subsection{The chaos of \pksf}


This paper presents 20-years of  MWL observations of the blazar \pksf, covering the period 2005–2024 and combining  $\gamma$-ray,  X-ray, and  optical-UV observations.
This long-term analysis provides a uniform view of the variability and spectral evolution of one of the most famous FSRQs. 
The analysis shown in this work provides a long-term view of the source’s emission across different energy regimes.

It is evident that on these long time-scales barely any strong relation between the bands exists. While the flux-flux plots indicate correlated activity at times between some bands, it is not uniform and, hence, correlations change from epoch to epoch. This also explains why older analyses, such as \cite{Castignanietal2017} and \cite{yuan+23}, found a DCCF value of $\sim0.5$-$0.6$ between HE and optical bands at 0\,d lag using data spanning roughly the first 8 and 10 years of \fermi\ operations, respectively, while in our data set the DCCF value is less than $0.4$ and not significant, in-line with results of \cite{amadorportes+24}. The DCCF becomes too diluted when mixing all the various events and epochs.

The variability is most pronounced in the HE \g-ray domain. This is shown by both the range of fluxes attained, which is more than 2 orders of magnitude, and the \fvar, which is generally higher than in any other band. The \fvar\ correlates with the flux state in this band. 
The flux distribution is compatible with a log-normal function. This is different from the results in \cite{kushwaha+16} who found that the HE band is described by a double-log-normal function with two neatly separated distributions, which is commonly interpreted as describing two different states \citep[as done for \pkst\ in, e.g.,][]{hess10_2155}. 
As described in detail in \cite{hess23}, the HE band exhibited a significant flux drop in July 2021, after which the fluxes and the variability have remained consistently low. Indeed, the fluxes since this date occupy only the lower parts of the flux distributions below the median of the total distribution. Additionally, the lowest three instances of \fvar\ have occurred between 2022 and 2024.

The optical flux range and the \fvar\ is surprisingly limited, as the optical domain is within the high-energy range of the synchrotron component. However, this domain is diluted by the brightness of the accretion disk, which provides a relatively stable (even though most likely not constant) lower flux level. This is corroborated by the flux distributions, which are fit well with double-log-normal distributions. In both the B and V bands, the two sub-distributions are rather different, allowing us to interpret one as being dominated by the accretion disk, while the other one at higher fluxes is dominated by the jet. This surmise is corroborated by the fact that the flux distributions since the flux drop in 2021~-- which effected the optical domain much like the HE band~-- is only filled within the potential accretion-disk-related sub-distributions. This is different from \cite{yuan+23}, who concluded that in their data set the jet always dominates the disk. However, the optical polarization data presented in \cite{hess23} and \cite{podjed+24} clearly show that the optical flux since the flux drop is unpolarized and thus most likely strongly disk-dominated.

The X-ray band is the least variable overall, but can exhibit orphan flares, where a significant flux rise is not met by any activity in the optical or HE bands. 
The low activity suggests that the radiative output in this energy band is driven by relatively-low-energy particles, for which the cooling time is long and variations are thus slower. It also suggests that variations in the entire SED are rarely driven by simple changes in the particle normalizations, as this would result in the same amplitude variations in all bands \citep[e.g.,][]{thiersen+24}; especially, if the normalization variations were dominating any cooling effect. Instead, the variations are more likely driven by intricate, energy-dependent processes resulting in, for instance, variations in the particles' spectral index.
Unlike the HE \g-ray and the optical bands, the X-ray band was apparently not influenced by the flux drop, as the flux since July 2021 still cover roughly the same flux range as the total distribution. However, the low number of observations makes a final interpretation tricky.

The spectral evolution in the X- and \g-ray bands does not follow any particular trend. The X-ray index is variable, but not in any obvious relationship with the flux. 
While the \g-ray spectrum can experience significant hardening, even to indices below $2.0$, this behaviour is not consistent throughout. This may be related to our time-binning of $4\,$d, as the observed flares at VHE \g\ rays \citep{magic17,hess+21,hess19ATel} are related to hard HE \g-ray spectra that we do not find here. Interestingly, the majority of hard spectral states in the HE band is found after the flux drop. This is in-line with the findings of \cite{hess23} that the flux drop was accompanied by a significant hardening of the spectrum.
\cite{dengjinag24} studied the spectral evolution of the $30\,$d-binned HE \g-ray light curve by cutting the energy range into 4 energy bins and then obtaining the flux ratios of the three higher bins to the lowest one. With this method, they found that the HE band displays a softer-when-brighter trend for low-flux states, and a hard-when-brighter trend for high-flux states on these long time scales. 

Naturally, some of these observations suffer from data completeness issues \citepalias[for a detailed discussion, see][]{WZ26}, as for instance the brightest \g-ray flare in the later parts of 2011 is not covered with \textit{Swift} data -- probably due to solar angle constraints -- or the significant X-ray and optical data gap(s) between 2017 and 2021. On the other hand, some flares have been covered in all bands, like the multiple-burst period in 2015, and here the individual bands reached their respective maximum flux of that year during different flares.

The findings above paint a very chaotic picture of the kinetic processes within the jet of \pksf. While at least a two-zone model has reasonably been established by \cite{hess23}, these authors separated the zones also by variability properties. They suggested that the secondary zone (producing the VHE and significant parts of the X-ray emission) was less variable than the primary zone (producing most of the HE and optical synchrotron flux). Following this separation, the primary region would have been responsible for most of the flares we see in Fig.~\ref{fig:1510_mwl_lc} and thus drive the (non-)correlations between the bands. However, reconciling these varying correlations within a single zone is challenging. A simple change in the particle distribution would influence especially the optical and HE bands in the same universal manner. As this has not been observed, parameters, such as the magnetic field or external photon fields, would have to vary in specific ways in order to remove the universality of the flux changes \citep[see, e.g.,][]{hess19}. Hence, a multi-zone ($N>2$) model seems a more plausible option, as the source region's parameters (magnetic field strength, external field fluxes) would naturally differ from region to region. Whether this implies multiple spatially-stationary zones that are active more or less simultaneously, or moving components interacting with, e.g., recollimation shocks \citep{hess+21,zacharias23}, requires detailed modelling, which is beyond the scope of this paper.

From the time-evolution of the flux and the \fvar, especially in the HE \g-ray band, it is clear that the jet has changed its character. The first obvious change took place between 2016 and 2017, after which the overall variability is less than what it was before [see Fig.~\ref{fig:1510_mwl_lc}(f)]. While there was still significant MWL activity after this \citep[e.g., the second VHE outburst in July 2019,][]{hess19ATel}, the HE \g\ rays had already reduced their variability considerably. This has become even more obvious after the flux drop in July 2021 \citep{hess23}, after which the overall HE variability is very low, and the flux range attained by both the HE \g\ rays and the optical fluxes is very limited. This slow reduction in flux and variability can also be interpreted as a slow increase in the angle between the jet and the line-of-sight leading to a reduction in the Doppler beaming \citep[e.g.,][]{dammando+19}. However, the rapidity of the 2021 flux drop does not suggest a gradual change at that point. Additionally, the large variations in the variability behaviour before 2017 also suggests that these variations were not induced by a change in the Doppler beaming alone, but requiring at least a mix of changes in Doppler beaming and other jet parameters as discussed before.

\subsection{Comparison between an FSRQ and an HBL}

In the studies presented here and in \citetalias{WZ26}, we employed the same analysis techniques allowing us to perform a unique comparison: An FSRQ, \pksf, with an HBL, \pkst. 
Here, we discuss their similarities and differences

In \pkst, the variability is most pronounced in X-rays, moderate in HE $\gamma$ rays, and weakest in the optical band, while the order in \pksf\ is that the highest variability occurs in the HE \g\ rays, is moderate in the optical domain, and weakest in the X-rays. Both sources are thus consistent with the expectation that higher energies in a given non-thermal SED component exhibit higher variability. This is underlined by both the range of fluxes attained in a given band, as well as the associated \fvar\ values.

In \pkst, all flux distributions are compatible with log-normal distributions, even though the optical one (low-energy part of the synchrotron component) is influenced by a change in baseline flux around 2009 after which the optical fluxes are consistently lower.
In \pksf, the optical flux distributions (B and V bands) are consistent with a double log-normal distribution, where one of the sub-distributions is dominated by the accretion disk, while the other one describes the dominance of jet high states.
Log-normality has been interpreted by some authors as a sign for multiplicative processes operating in blazar jets \citep[e.g.,][]{Uttley2005,hess17_2155,Shah}, even though others disagree with this conclusion \citep[e.g.,][]{BiteauGiebels12,Scargle20}.

In terms of flux-flux correlations, the results between the two blazars are comparable. Except for the X-ray-\g-ray case in \pksf, there are positive correlations between the various bands. However, there is never a unique correlation, and various tracks are visible suggesting variability with varying degrees of correlations from event to event. This is further highlighted by the DCCF analysis, which does not find any global correlation at any lag studied. The latter is a sign that any lag from one flare is washed out by differing lags from other flares.
Hence, each event operates differently in terms of particle acceleration, and other involved parameters of the flaring region (e.g., magnetic field or seed photon fields).

The latter fact is further underlined by the spectral variations in each source.
\pkst\ shows a clear harder-when-brighter trend in the X-ray band, that is the high-energy part of the synchrotron component. Interestingly, the slope of this trend evolves from year to year, reflecting changes in particle acceleration efficiency. This is remarkably different from other HBL sources, such as Mrk\,421 and Mrk\,501, where the harder-when-brighter trend is rather uniform \citep[e.g.,][]{Taylor+26}.
The corresponding spectral range (high-energy part of the synchrotron component) in \pksf\ would be the optical domain. However, the presence of the accretion disk does not allow us to study the spectral evolution of the jet in this band. However, the HE \g-ray band, which represents the high-energy part of the high-energy SED component, does not exhibit any persistent spectral--flux relation. The \g-ray spectrum is variable, but there is no global trend. While some hardening events ($\Gamma<2$, $\alpha<2$) are associated with major MWL flares \citep[e.g.,][]{hess+21}, they do not correlated with the HE \g-ray flux. Since the flux drop in July 2021, the HE \g-ray spectrum is persistently harder with an index close to $2$ or below.
In either source, the low-energy domain of the high-energy SED component (HE \g\ rays in \pkst\ and X-rays in \pksf) does not show any spectral trends.

A challenging aspect of blazar variability are so-called ``orphan events'', where the variability is limited to a single band. While we only consider three energy bands and thus cannot rule out correlations with other energy bands (such as the VHE \g-ray band), it is still worth looking for these events in our data sets. In \pkst, we find two such occasions: an orphan optical flare in 2016, and a very strong X-ray flare in 2024 (actually, the brightest X-ray state in our data set). In the latter case, the lack of counterparts is also underlined by the \fvar, which were low in both the HE and optical bands for that year. In this case, only the high-energy part of the synchrotron component was effected, while the low-energy parts of the SED components were almost steady. 
While the orphan optical event in 2016 is evident \citep[for a more detailed light curve, see][their Fig.~1]{2155_monit2015}, it is not as obvious from the variability aspects as the 2024 event. While the \fvar\ is relatively low for both the X-ray and HE \g-ray bands (below $0.5$), they are still higher than the optical \fvar. This may be related to the fact that the optical variations proceeded relatively slowly also in this case.

In \pksf, there is just one clear orphan event in our data set: an X-ray flare, which took place in 2017. In this source, this of course means that the low-energy part of the high-energy component was flaring, while the high-energy parts of each SED component remained calm. Indeed, the \g-ray \fvar\ is close to $0.4$, which is roughly the minimum \fvar\ for this band (before 2021), while the optical \fvar\ is below the X-ray one. We also would like to point out a few more events, even though they are not orphan. In 2010, both the X-ray and HE \g\ rays varied, but the optical domain was not variable at all, as also evidenced by the \fvar. Similarly, during the 2011-2012 flaring episode, we see a similar evolution, even though the brightest \g-ray peak is unfortunately not covered with \textit{Swift} probably due to solar cut-outs. In these cases, the entire high-energy SED component flared, while the optical flux was steady. In the 2022 and 2024 data sets, the only detectable variability is in the X-ray band (as also evident from the \fvar\ values in Fig.~\ref{fig:1510_mwl_lc}(f)), while in 2023 only the optical band shows mild variability. 

Whether or not the orphan events may have counterparts in energy bands not covered in our study, they are still puzzling entities. Within a compact flaring region, one naively expects relatively close correlations between the various bands. The reason is that, even if the initial flare took place at higher energies, the responsible particles should eventually cool down and also cause a flare in lower energy bands. If this does not happen, particles must stop radiating before reaching lower energies, be deflected in a direction not well visible to us, 
or the emission region exhibits very fine-tuned parameter variations to avoid the emission of a large number of low-energy photons \citep[such as the model applied to the 2015 flare of 3C\,279;][]{hess19}. Multi-zone and extended-jet models also provide a potential basis for orphan flares, as the location of the flaring region within the jet \citep{wang+22_orphanmodel} may play a role along with the requirement that the flux must rise above the quiescent flux of all other zones in the jet \citep{potter17_orphanmodel}. 

A flare only at the low-energy parts of the SED components, as seen in \pksf, implies an overabundance of low-energy particles, while the high-energy particles remain at their normal quantities. 
Within a simple one-zone model, an increase in low-energy particles may be achieved by lowering the break energy in the particle distribution compared to the nominal state. This could be achieved by increasing cooling and/or by reducing particle escape. If otherwise the processes remain similar (e.g., injection profile), the particle distribution and, in turn, the SED, may not change at higher energies, while the higher number of particles at lower energies will cause a rise in the corresponding SED range.
The orphan optical flare in \pkst, which implies variability only in the low-energy synchrotron component, would additionally require significant fine tuning in order to avoid the inevitable inverse-Compton flare. Naturally, if the high-energy SED component would stem from hadronic processes, the protons may not react to a change in the source parameters as much as the electrons. 

The flux drop in \pksf\ in July 2021 revealed that this source used to have at least 2 emission zones, even though the second zone appears to be non-variable \citep{hess23}. While there is no clear proof yet for more than one zone in \pkst, the curious spectral hardening at hard X-rays observed in April 2012 \citepalias{WZ26} could be interpreted in this regard. However, the feature in \pkst\ is variable. Hence, the jet of \pkst\ would differ here from the jet in \pksf. Further explanations for the X-ray hardening in \pkst\ include, for instance, signs of hadronically-induced emission.


The 20-year-long light curves presented here and in \citetalias{WZ26} should provide us with a clear view of long-term evolution of the jets in these two sources. Indeed, the slow decay of the \g-ray variability in \pksf\ is such an interesting long-term trend. The flux drop in July 2021, which has been accompanied by a similar drop in the optical, was, however, rather sudden. The aforementioned multi-zone model can describe the data, but the cause of the drop remains unknown. In \pkst, we see a long-term evolution only in the optical bands, which exhibit a change in their baseline flux around 2009; however, the variability properties have not changed. If, for instance, the HE \g-rays also exhibited such a flux drop is, unfortunately, impossible to say, as the observations only began in late 2008 shortly before the drop. Our data set also does not reveal how sudden the flux drop was in \pkst\ due to a lack of observations with \textit{Swift}. 

Potential causes for long-term changes can be a (time- and spatially-dependent) bending or lateral motion of the jet. The induced changes in the angle between the jet direction and the line-of-sight would result in a change in Doppler boosting that can have dramatic consequences on the flux \citep{dammando+19}. Coupled with the multi-zone interpretation of the stochastic variations, the bending could lead to zones being effected in different ways resulting in non-uniform effects on the observed energy bands. This is one of the interpretations given by \cite{hess23} for the flux drop in \pksf. There are several potential causes for such a lateral motion of the jet, such as kink instabilities in the jet \citep[e.g.,][]{tchekhovskoy2016three}, a precession of the black hole itself (perhaps induced by a companion black hole) resulting in a helical jet \citep[e.g.,][]{1993A&A...278..391S}, interactions of the jet with the non-isotropic flow of the surrounding medium \citep[e.g.,][]{zacharias+17}, and others. Further, consistent monitoring is needed to check if the sources exhibit other long-term evolutions, or if the fluxes return to previous levels.

%
%
\section{Summary} \label{sec:sum}
We have presented a 20-year multi-wavelength variability study of the blazar PKS\,1510-089, using \textit{Fermi}-LAT  data, as well as \textit{Swift} XRT and UVOT observations obtained between 2005 and 2024.
The main results of this work are summarized as follows:

\begin{itemize}
\item The long-term multiwavelength data reveal strong flux and spectral variability in PKS~1510$-$089 across the $\gamma$-ray, X-ray, and optical bands over the past two decades.

\item The flux distribution is energy-dependent: the HE \g-ray and X-ray bands follow a log-normal distribution, while the optical bands are compatible with double-log-normal distributions. The latter may be attributed to the disk and jet contributions dominating to various degrees at different times. In any case, the widths of the MWL distributions reflects the general trend that fluxes attain a broader range of values the higher the energy band is located within a given SED component.

\item No significant flux–flux correlations are observed between the $\gamma$-ray, X-ray, and optical bands. The discrete cross-correlation analysis also reveals no persistent interband lag. 
This suggests different physical conditions in the jet from one flaring event to the next, as any strong correlation during one event is washed out by the others.

\item The $\gamma$-ray spectra show strong flux variability but only weak spectral changes, with photon indices occasionally reducing to below~2.0. 
No correlation with specific events is observed. However, the majority of these hard states occurs since the flux drop in July 2021.

\item The X-ray spectra are significantly variable in flux and photon index but do not show any specific relation, suggesting complex or variable acceleration and cooling conditions within the jet.

\item The dramatic flux drop observed in July~2021, accompanied by the disappearance of optical polarization and a decline in optical and HE~$\gamma$-ray fluxes~-- while X-ray and VHE~$\gamma$-ray emission remained nearly constant~-- provides strong evidence for the presence of multiple, spatially separated emission zones \citep{hess23}.

\item The fractional variability amplitude, \fvar, is largest in the $\gamma$-ray band, moderate in the optical, and lowest in X~rays. This trend 
is in-line with the expectations that high-energy parts of an SED component are more variable than low-energy parts. Since the aforementioned flux drop, the \g-ray and optical variability is much reduced.

\item The variability amplitude correlates with the average flux level in the $\gamma$-ray and optical bands, while the X-ray band shows no such dependence, suggesting that different mechanisms drive variability at different energies.

\item The absence of persistent interband correlations, as well as the band-dependent flux distributions 
argue against a simple one-zone leptonic scenario describing the entire presented data set. Instead, this seems to favour a multi-zone or stratified jet model in which distinct regions contribute independently to the total emission.

\end{itemize}
We point out that data completeness (and other limitations), as discussed in detail in \citetalias{WZ26}, also plague the data sets of \pksf. However, the long duration of our data sets enables robust conclusions.

Gathering the results of this paper, as well as of \citetalias{WZ26}, the comparison between an FSRQ and an HBL yields the following summary:

\begin{itemize}
\item In both sources, the variability is higher for higher-energy bands within a given SED component. This is underlined by both the flux range and the \fvar\ values.

\item Interband variations never follow a single track suggesting that each variation (of small and large amplitude) behaves differently. This is supported by the lack of any significant cross-correlation using the full data sets.

\item While the high-energy synchrotron component in \pkst\ shows harder-when-brighter behaviour, no such indication is found in any band in \pksf. However, the presence of the accretion disk in \pksf, as well as general data limitations may inhibit finding such relations.

\item Both sources exhibit orphan flares, where only one observed band varies, while the others remain nearly constant. While data limitations also play a role here, it is still interesting to point out that these events seemingly can take place in any energy band irrespective of ``location'' in the SED component.

\item The few truely long-term changes observed in both sources (e.g., flux and variability drops) might be related to long-term changes in the jets themselves. Coupling a time- and spatially dependent lateral motion of the jet with the multi-zone interpretation may explain, why these changes do not happen in all observed bands.


\end{itemize}

The long-term studies of these two sources, \pkst\ and \pksf, which represent well the FSRQ and HBL blazar classes, demonstrate that even over decades blazar variability is not a universal or predictable feature. In other words, even the variability is variable. Nevertheless, despite their many differences, the two sources have also shown similarities in their behaviour. This underlines the -- seemingly obvious, but nevertheless important -- fact that the jet physics are comparable between the different source classes.

%
%
\section{Acknowledgement}
The authors thank the anonymous referee for a constructive report that helped improving the manuscript.
The authors are grateful for stimulating discussions with G.L.~Taylor, S.~Wagner, J.-P.~Lenain and D.~Dorner.
The authors gratefully acknowledge the Polish high-performance computing infrastructure PLGrid (HPC Center: ACK Cyfronet AGH) for providing computer facilities and support within computational grant no. \text{PLG/2024/017925}. %

\section{Author contributions}
M.~Zacharias (Conceptualization, Methodology, Writing – original draft),
A.~Wierzcholska (Conceptualization, Data curation, Formal analysis, Writing – original draft)

\section{Funding sources}
The project on which this report is based was funded by the Bundesministerium für Bildung und Forschung (BMBF, Ministry of Education and Research). Responsibility for the content of this publication lies with the author. 
The project is co-financed by the Polish National Agency for Academic Exchange. %
This work is funded in parts by the Deutsche Forschungsgemeinschaft (DFG, German Research Foundation) -- project number 460248186 (PUNCH4NFDI). %
%

%
%

%

\bibliographystyle{elsarticle-harv} 
\bibliography{references}
\end{document}